\documentclass[journal]{IEEEtran}

\usepackage{setspace}
\usepackage{color,soul}
\usepackage{bibentry}
\usepackage{float}
\usepackage{multirow}
\usepackage{algorithm}
\usepackage{algpseudocode}
\usepackage{mathtools, cuted}
\usepackage{cite}
\usepackage{hhline}
\usepackage{graphicx,bm}
\usepackage[caption=false]{subfig}
\usepackage{setspace}
\usepackage{stfloats}
\usepackage{color}
\usepackage{cite}
\usepackage{amsmath, amsfonts, amssymb}
\usepackage{epstopdf}
\usepackage{commath}
\usepackage{verbatim}
\graphicspath{{./fig/}}

\begin{document}

\title{Digital Twinning in Smart Grid Networks: Interplay, Resource Allocation and Use Cases}
%\title{Digital Twin-Empowered Smart Grid: A Radio Resource Allocation Perspective}
\author{Abdullah Othman, Georges Kaddoum, Joao V. C. Evangelista, Minh Au and Basile L. Agba
\thanks{A. Othman and G. Kaddoum are with Electrical Engineering Department, École de Technologie Suprieure, Université du Québec, Montréal, QC H3C 1K3, Canada (e-mail: abdullah.othman.1@ens.etsmtl.ca; georges.kaddoum@etsmtl.ca).}
\thanks{J.V.C Evangelista is with Ericsson Canada, Ottawa (e-mail: joao.victor.de.carvalho.evangelista@ericsson.com)}
\thanks{M. Au and B.L. Agba are with the Hydro-Quebec Research Institute (IREQ), Varennes, QC, Canada (e-mail: au.minh2@hydroquebec.com; agba.basilel@hydroquebec.com)}}
\maketitle

\begin{abstract}
Motivated by climate change, increasing industrialization and energy reliability concerns,
the smart grid is set to revolutionize traditional power systems. Moreover, the exponential annual rise in number of grid-connected users and emerging key players e.g. electric vehicles
strain the limited radio resources, which stresses the need for novel and scalable resource management techniques. 
Digital twin is a cutting-edge virtualization
technology that has shown great potential by offering solutions for inherent bottlenecks in
traditional wireless networks.
In this article, we set the stage for various roles digital twinning
can fulfill by optimizing congested radio resources in a
proactive and resilient smart grid. Digital twins can help smart grid networks through real-time monitoring, advanced precise modeling and efficient radio resource allocation for normal operations and service restoration following unexpected events.  However, reliable real-time communications, intricate abstraction abilities, interoperability with other smart grid technologies, robust computing capabilities and resilient security schemes are some open challenges for future work on digital twins.

%Comparison with other benchmarking algorithms demonstrates that the proposed method shows more robust and accurate performance. 
\end{abstract}
% \fi
\IEEEpeerreviewmaketitle

\section{Introduction}% This is hoped to be the new intro
Events with low probability and wide-scale impact, such as earthquakes and hurricanes, can cause significant outages. Data reveals that between $2000$ and $2016$, over $1,500$ major outages occurred in the continental United States, where 
a major outage implies that at least $50,000$ customers were affected or over $300$ MW of power was lost \cite{SayantiMukherjeeRoshanakNateghi2018}. Factors such as cyber-attacks, climate change and overpopulation further highlight the need for new smart grids (SGs) that can provide reliable power output and integrate more renewable energy resources while ensuring resiliency against unexpected disasters. The $2015$ cyber-attack on the Ukrainian power grid is a prime example of a challenging man-made event that left around $230,000$ people without power for nearly six hours \cite{Whitehead2017}. 

Moreover, an SG should adapt to harsh conditions and scale up to address new challenges introduced by dynamic loads due to electric vehicle (EV) and intelligent transportation use. Given that the spectrum is limited in size and congested, and an exponential increase in the number of new users connected to power grids, optimized radio resource allocation using wireless communication technologies remains a significant challenge for network management, where efficient utilization of network resources is the holy grail.

However, SG environments differ widely in their network requirements. While some applications demand high data rates and flexible latency constraints, others, such as mission-critical SG communications, have stringent latency requirements with variable data rates.
%Simply put, resource allocation is defined as transmit strategy selection for network users subject to transmission power constraints \cite{Bjornson2012}. Typically, resource allocation considers one or more user satisfaction metrics known as quality-of-service (QoS) indicators.
Traditional network management presents numerous drawbacks that make it inefficient for pervasive and resource-demanding SG applications. Network resources are dedicated in fixed and inflexible ways. Recently, virtualization methods that use network slicing tools have unlocked the research landscape's potential for creative network resource management. Decoupling the physical and resource optimization layers ensures more autonomy and efficiency when utilizing limited network resources, e.g. power and bandwidth. 
In particular, SGs can benefit from the 6G revolution by using \textit{digital twins} (DTs) to manage their communication networks more effectively. DT is a virtual mapping technology that enables creative resource allocation in future wireless networks by integrating artificial intelligence (AI) and other modern technologies \cite{Bariah2022a}.

While no full-fledged DT products currently exist \cite{Minerva2020}, in this article, we explore in depth the interplay between digital twinning and the SG, and in particular how DT can provide intricate modeling and detailed resource allocation for different SG networking applications. Furthermore, we develop a key generalized architecture for DTs in operational SGs, whose purpose is to orchestrate simultaneous real-time monitoring and resilience against unforeseen circumstances. Moreover, we discuss communication technologies that have the potential to support DT schemes and how the connections between twins can enhance an SG's ability to handle novel challenges and adapt to innovative changes in renewable resource and EV scenarios. Finally, we lay out future challenges and a road map for the innovations needed to handle these challenges.

%\textit{Digital twin} (DT) is an emerging digital mapping technology that provides a solution for
%intelligent resource allocation in future networks by creating a real-time model simulated from
%physical objects. More specifically, DT implements a two-way closed-loop feedback process
%of dynamic information between the physical and virtual entities. More than only collection
%of real-time data about physical entities, DT also implements control to change state of these
%physical entities. Hence, DT optimizes overall network quality-of-service (QoS) by dynamically enabling real-time
%monitoring and scheduling resources in dynamic scenarios.

% Brief sentence on resource allocation and its importance

%A microgrid (MG) is a self-sufficient local energy grid that uses renewable energy sources (RES)s to generate and distribute power to the UEs. However, it could also trade energy with other MGS and the central grid, whenever needed.

%In NB-IoT, we assume a path-loss model where the attenuation is defined as x^(-α), where x is the propagation distance from the BS and α is the path-loss exponent.

%\begin{figure}
%	\centering
%	\includegraphics[width=9cm]{DPSO_AMPSO_resized.eps}\\
%	\caption{AMPSO and DPSO flowchart}
%	\label{fig:ampso_dpso}
%\end{figure}

\section{Digital Twins to Optimize Wireless Resources}

SG wireless networks are envisioned to serve a staggering number of users, where small and inexpensive internet of things (IoT) sensors generate the bulk of transmitted data. Hence, it is necessary to build scalable and efficient methods to manage the limited radio spectrum and transmit the data generated while considering the inherent delay and power constraints of each data type. 
Furthermore, SG communications are hindered by the prevalence of impulsive noise from the harsh environment in substations \cite{Agba2019}.  Impulsive noise is characterized by strong radio frequency interference with broad spectral density due to the high-voltage equipment that is present, which significantly deteriorates quality of service. %This poses another challenge to DT-based resource allocation in grid environments and must be addressed in any proposed schemes. 

%Conventional optimization methods are inappropriate in this case because the additive white Gaussian noise (AWGN) assumption is not realistic in SG environments.

%Digital twins open a new world by which radio resource optimization takes up a whole new level. By breaking up the world into two spheres, which are independent
%but nevertheless mirror each other, we delve into a whole world of new possibilities. The digital twin acts on the data collected from the physical twin. Resource optimization
%in the digital twin goes into a host of activities, including diagnostics, prediction, comparison with other physical systems under surveillance by other DTs etc.

%DT modeling offers significant reductions in computation and storage delays by only recording changes in the model parameters as opposed to storing data of physical devices. Moreover, DT offers several benefits for IoT and industrial IoT (IIoT) networks.
%For example, DT is a digital mirror of the physical system, such that the two entities grow in sync with each other during the product lifetime \cite{Nguyen2021}. In particular, DT enhances preventive maintenance frameworks, supports creative business models, optimizes product efficiency and sustainability.
%Through bi-directional communication between the digital and physical world, DT could enable real-time engineering decisions such as resource management in  telecommunications.

\subsection{New radio resource management world}

DTs convey a holistic representation to model complex systems with heterogeneous data types. 
Each DT would store the data of its physical twin and monitor it to record any parameter changes. 
Hence, data is constantly gathered, processed, and shared to provide resource allocation solutions to the physical plane. 
DTs further transform resource optimization by making cyber twins completely autonomous virtual agents that work independently from their physical twins. Novel resource optimization schemes are created through an endless process of data updates, model building, parameter estimation, precision enhancement, and innovative optimization using various tools at its disposal. 

DTs slice the networks using a customized set of features where each DT has specific functions and provide a specialized set of services.
Monitoring twins describe the current physical state of the twin, requiring real-time communication and a locked feedback loop to reflect simultaneous changes. Analysis/simulation twins describe the hypothetical state of the physical twin under various scenarios. These scenarios can be used to improve the system's state or to predict a bad state given an unexpected event or series of events. The simulation twins then develop strategies to either achieve the plausible outcome or reduce the effects of the negative outcome by manipulating the wireless network and employing cutting-edge technologies.

For example, a local metering network composed of a few smart meters to measure energy consumption requires a simple network of monitoring twins that clone and survey the meters' state and properties in real-time. On the other hand, a heterogeneous multi-purpose internet of vehicles (IoV) network consisting of tens of EVs, roadside units, and base stations (BSs) should encompass monitoring twins of all the components to abstract all their details and capture the intricate relationships between the objects, and simulation twins that work towards achieving various objectives, e.g. collision avoidance, autonomous driving, entertainment purposes, and pedestrian safety. This level of complexity could make DT network implementation prohibitively complicated.  
	We classify the main tasks the SG-DTs are envisioned to undertake as follows:
	\begin{enumerate}
	\item Real-time monitoring and routine management of physical twins.
	\item Short-term model sharing from monitoring DTs to analysis DTs.
	\item Long-term model propagation from analysis DTs to monitoring DTs.
	\item Emergency response planning and coordination among DTs and plan implementation by the physical twins.
	\end{enumerate}

Task $1$ takes place between the digital and physical planes, and its requirements differ depending on the SG use case. Hence, adaptive resource management solutions are needed. Tasks $2$ and $3$ occur in the digital plane and hence require different resource management tools than those dedicated to Task $1$. On the other hand, Task $4$ requires resilient and ultra-reliable low-latency communications.
Hence, several open questions require further investigation in DT-enabled resource optimization \cite{Bjornson2012}. 

For example, which communication technology best suits the network requirements of SG applications, and how can channel access, user adaptation and packet scheduling in the communication channel be facilitated? 
DT-enabled resource allocation requires a flexible scheme that supports various SG applications and functions. For example, when a high data rate is needed, orthogonality and synchronization can be maintained using fixed grant-based schemes at the expense of longer delay. However, random access resource allocation methods that offer high spectral efficiency are more suitable for massive IoT networks with sporadic traffic.

\subsection{AI-assisted resource optimization}

DT applications will be driven by big data collected from ubiquitous IoT sensors in different SG environments. Since the network's size and computational complexity keep increasing, a scalable approach is needed. 
Recent advancements in AI solutions show its potential to empower DT-based resource allocation applications. More specifically, machine learning (ML) tools are suited for DT scenarios in which labeled and unlabeled datasets are available for supervised and unsupervised learning, respectively.
However, data type selection, expert labeling, and proper pre-processing present significant challenges for data-intensive DTs \cite{Rathore2021}. 

Synthetic data generation is an alternative for building the initial model at the DTs. For example, generative adversarial networks can be coupled with power system analysis platforms to create realistic datasets relevant for DT set-up.
Moreover, semantic reasoning can be leveraged to intertwine the synthetic data and processed actual data for model training \cite{Letaief2019}. 
DTs can then use co-simulation platforms to imitate different grid scenarios and test ML and deep learning (DL) algorithms before their implementations in the physical realm~\cite{Bian2015}.
On the other hand, reinforcement learning (RL)-based algorithms are suitable for cases in which the datasets are unavailable, but a virtual environment could be created to mirror the physical twin. 
Realistic virtual environments can be designed to mimic actual scenarios in vehicular networks, and RL-based agents can be leveraged to intelligently emulate physical twins and create  novel resource allocation methods.
Distributed resource allocation is pertinent to massive SG networks, where RL-based algorithms are scaled to match the system's overall complexity. 
However, DL and artificial neural networks have been used as a black box without analyzing the driving forces behind their mechanism, which limits their success \cite{Bariah2022}. 

Mathematical formulations and model-driven tools can help us better understand, customize and utilize these AI methods. Therefore, joint examination of DL, ML, and mathematical-based formulations can be leveraged to reap the benefits of new frontiers for DT resource optimization. 
Contrary to traditional offline learning systems, DTs can start building their model using offline learning, then continue maturing over their life span using online learning through new experiences. Feedback on digital twinning performance comes from users' practical experience with physical twins. This feedback would help to remodel the DTs' operations and adjust their learning process.

%\hl{<AI and DTs interplay>
%	- How AI need to be customized for DT 
%	Supervised, unsupervised and reinforcement. DT collecting the data creates huge datasets to be used.
%	RL saves huge memory allocations by trying to model the exact replica of physical underlying env.}

%AI techniques which have progressed greatly recently can be exploited to run on big data and find trends and solutions to chronic SG problems such as fault-reaction and outage 
%recovery, service deterioration in sub-urban/rural areas, and SG cyber-security.

%In other words, it will be continuous, proactive, independent and more creative.

%DT can enable virtual building and testing for 5G use cases such as mMTC and mission critical IoT. 
%Furthermore, virtual launching, feedback and reviewing can be done within paradigm of DT.

%\hl{Layer 1 Level2 deals with resiliency during impact to alleviate the pressure on available resources, and restoring normal operation in 
%lower time and with minimum effects. This level works on donating available resources from other units to affected unit, and
%pooling other layer2 resources to relieve pressure on scarce resources. This is especially important for mission-critical events 
%and resources.}

\subsection{Imaginative problem-formulation and problem-solving}

Preparing the grid for unexpected events is a fundamental characteristic of simulation twins. 
This drives the development of alternative scenarios, and running of diagnostic analyses to discover new patterns for different SG use cases. Planning can then be modified accordingly. 
When analysis twins consider what-if scenarios, they could use insights from past events to foresee likely future circumstances and study the best courses of action to take in such scenarios. This analysis of hypothetical events could lessen the magnitude of grid accidents and their effects on the overall power system and customers.

Resource management in the digital sphere must support self-management and self-correction so that any network layout changes ---which are frequent in SG--- can be adapted to for maintaining optimal performance. Simulation twins can instantiate such scenarios to run their AI tests and analyses and produce an adaptive set of optimized formulas for the SG. 
Distributed AI can learn solutions to complex problems formulated by the DT-empowered processes.
Simulating twins would be the centerpiece of trial and error in the digital plane to solve problems that physical twins may face. Those twins can ensure that different cutting-edge technologies, such as reconfigurable intelligent surfaces (RISs) and blockchain, are used to solve various problems in SG environments, e.g. impulsive noise and reliable energy trading. 

For example, DTs could simulate RIS placement strategies to create virtual line-of-sight links and mitigate impulsive noise \cite{Padhan2021}. Microgrid monitoring twins collect information about the state of distributed energy resources' (DERs) output, weather, and market information using many parameters, such as
	resource availability, energy cost, and trade history. Simulation twins use this data to design secure blockchain energy trading strategies and stabilize the grid by scheduling power transfer to the SG from local microgrids and vice versa. The virtual power plant concept can be empowered using monitoring and simulation digital twins working together to inspect the state of grid components, analyze various situations and take corrective measures when needed.
Moreover, in IoV scenarios, the EVs fight for
access to the congested spectrum and finite radio resources,
and these competitions must be replicated in the corresponding
monitoring twins. It is possible to find game-theoretic models and their corresponding
Nash equilibrium points for situations in
which multiple analysis twins are modeled as players in cooperative or
non-cooperative games that have similar or conflicting objectives
that mirror a real situation, e.g. several EVs coordinating their
charging or discharging schedules with rush-hour pressure on
the SG.

\begin{table}[]
	\footnotesize
	
		\centering
		{\caption{Network Requirements and Expected DT Model Complexity for Some SG Applications. \label{Apps_reqs}}} % Contrainte manuelle de la largeur de la légende
		\begin{tabular}{|l|l|l|l|}
			\hline
			\textbf{Application}                                               & \textbf{Data Rate} & \textbf{Latency} & \textbf{Modeling} \\ \hline
			Local Metering Network                                                                & Low         & Variable   & Low      \\ \hline
			Asset Management                                                                & Medium             & Low   &   Low     \\ \hline
			Demand-Side Management                                                                & Variable         & Low     &    Medium   \\ \hline
			Distribution Automation                                                                 & Variable        & Ultra-Low &   High      \\ \hline
			Renewable Energy Resources                                                               & Variable        & Low &  High \\ \hline
			Intelligent Transportation & High            & Variable & High  \\ \hline
			Outage Management                                                                 & Medium             & Low     &     Medium   \\ \hline
			Substation Automation                                                                 & Variable        & Ultra-Low     & Medium \\ \hline
			Transmission Line Monitoring                                                                & Medium         & Low     &   Low    \\ \hline
			Wide Area Situation Awareness & High & Ultra-Low & High \\ \hline
	\end{tabular}
\end{table}
\normalsize

\section{Structure of Digital Twins in a Smart Grid}

\subsection{Twins of networks}

SG communication networks can generally be divided into home area networks (HANs), neighborhood area networks (NANs), and wide-area networks (WANs). HANs are short-range networks that cover an area up to $100$~m~$\times$~$100$~m, and serve applications related to energy consumption with low data rates, up to $10$~kbps. NANs are related to distribution-level applications with a larger coverage area, up to $10$~km~$\times$~$10$~km, and higher data rates, up to $10$~Mbps. WANs serve the backbone communication networks that support power generation and transmission. Their applications require data rates between $10$~Mbps and $1$~Gbps, and a coverage area of $100$~km~$\times$~$100$~km. 

Table~\ref{Apps_reqs} presents the data rate and latency requirements for several SG applications, with demands varying greatly from one use case to the next \cite{Ghorbanian2019}. Table~\ref{Apps_reqs} also shows the expected level of model complexity of DT networks when they are initially deployed to serve the SG applications.
Therefore, DT resource allocation tasks should respond to the specific needs of SG networks. Small-size HANs might require less computationally complex solutions, 
in which optimization could involve few variables. Resource allocation solutions for these types of networks could therefore be computed locally, and the DTs could be set up close to the physical twins to reduce transmission delay.
	
However, for medium-range NANs and long-range WANs, resource allocation tasks could involve many more variables and hence be too complex to be accommodated locally.
Offloading computational requests to nearby edge, fog, or cloud centers can help DTs meet their load requirements. Nevertheless, the incorporation of DTs in the development of such complex networks should be done gradually and planned over some time. The planning stages for such large networks should include setting up the skeleton of DTs to start data collection. Over time, the models' granularity would become finer as analysis DTs perform their functions with AI tools and share their insights with monitoring DTs, and vice versa. The various DT models would then be expected to start fitting the data gathered, and the DTs would be able to provide the physical twins with optimized radio resource allocation service.

\begin{figure}
	\centering
	\includegraphics[width=9cm]{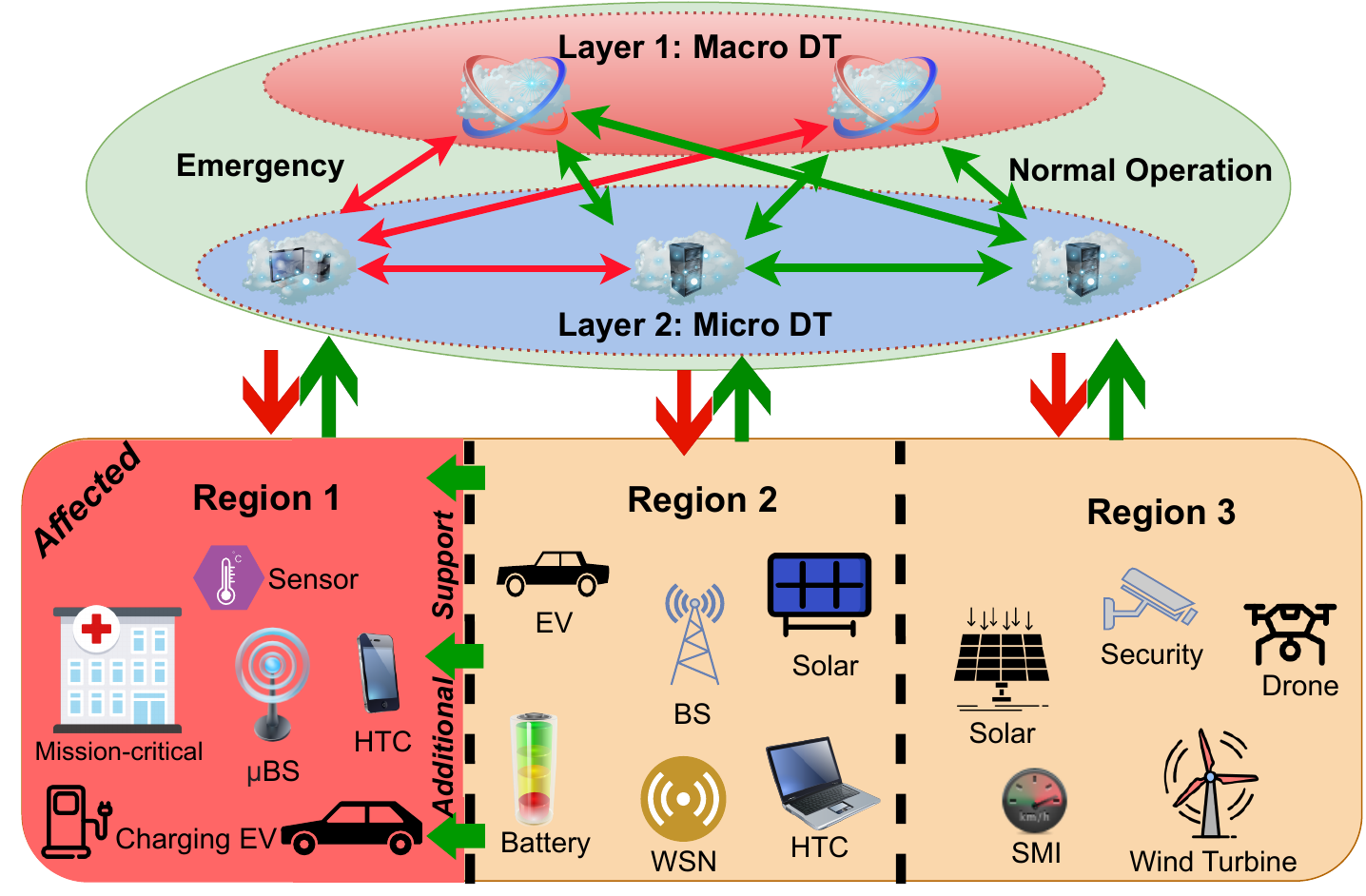}\\
	\caption{Two-layered Architecture for DT-based SG, where local DTs monitor their respective regions.  }
	\label{fig:DT_architect}
\end{figure}

% Macro and micro DTs coordinate their efforts to provide support for incident-affected region $1$.

\subsection{Networks of twins}

Future DT-monitored SGs require a multi-level structure of hierarchical DTs that collectively cooperate to ensure SG's proper functioning and safety. This structure would divide the DTs in their roles into two general levels, namely, micro and macro DTs. 
Fig.~\ref{fig:DT_architect} presents an example of a DT-supervised grid with the micro-DTs supervising different regions. In this architecture, tasks are divided, and the micro-DTs engage in concurrent mirroring with their respective physical twins. They can be deployed to local regions such as an urban area, a generation plant, or a residential microgrid. Such DTs would communicate using real-time communication technologies to ensure they have up-to-date data and that the model imitates the current state of the physical system. 

Macro DTs, on the other hand, could operate on a broader grid-level scale, supervising several micro DTs. They need not relate to the physical infrastructure directly but to and through the corresponding micro-DTs. Their task is to form an overall representative model that describes the general state of the underlying infrastructure. These DTs are meant to forecast events, ensure the health of the overall system, and orchestrate the response to unexpected but disastrous failures. In addition, they would support local DTs in the event of an emergency by helping with service recovery and operation restoration, and providing additional resources to impacted zones. 

SGs consist of various complex sub-systems that must operate together in perfect synchronization, such as generation plants, transmission lines, and substations.
Hence, correspondent digital replicas should include multiple interconnected DTs that survey the properties of different parts.
The International
Organization for Standardization (ISO) and the International
Electrotechnical Commission are working to develop DT-related
standards. Standard ISO~$23247$ provides general guidelines to build and represent DTs in manufacturing settings.
With new modeling and homogenized digital representations, DTs can abstract industrial IoT (IIoT) networks that have massive user distributions and are subject to stringent data requirements. These test beds help SG operators employ adaptive cross-layer algorithms to plan intelligent predictive maintenance and early fault detection. Moreover, cyber twins can perform long-term grid planning and event forecasting to prepare the grid for unforeseen situations. 

For example, within a substation communication network, distributed nodes such as intelligent electronic devices record measurements taken by potential and current transformers and report all parameter changes to the substation central controller. This process is mimicked at the monitoring DTs, and the collected data is analyzed to investigate it, anticipate events, and provide early detection and correction mechanisms.

%Talk about integrating DERs and matching theory to link physical and cyber twins
%An especially important feature of DT is when multiple interconnected DTs are considered. This notion could open the possibilities for a digital future where a plane of DTs might gather data, exchange useful information, process them and provide solutions to the user plane that enhances efficiency in industry, automation and telecommunication fields to a great extent.

%\hl{- How SG apps require DT RRM to solve inherent problems. Use virtual power plant, diagnostics and fault detection
%and correction.
%- DT-based RRM can use forecasting of renewables generation and load dynamics to control the levels of power generation to determine
%how much output to be used from the solar cells, to the fossil fuels to the energy storage. This would limit the carbon footprint
%and improve the participation of renewable energy resources in the future smart grids. DTs can also work to better integrate the renewables
%by linking different factors such as weather, time of the year, load historic and forecasts and economic implications to better predict
%and manage the often unpredictable renewables output. Deployment of DTs on wide areas and for long time frames can generate gigantic data
%sizes, and running AI tools on these real-data would incur significant improvements in prediction and analytics domains.}

\begin{figure*}
	\centering
	\includegraphics[width=18cm]{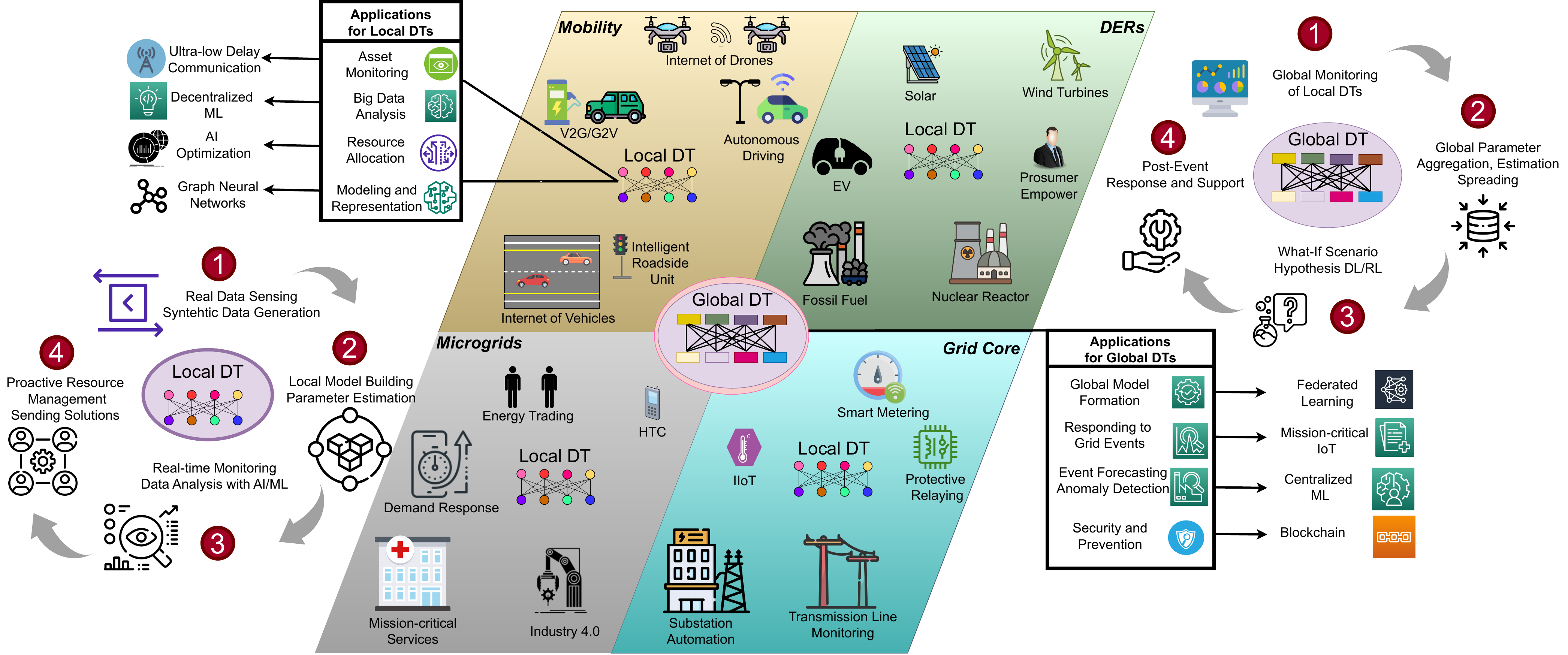}\\
	\caption{DT-empowered SG, with different applications for local and global DTs. The set of objectives and enabling technologies for each DT are presented.} 
	\label{fig:DT_lifecycle}
\end{figure*}

\subsection{Inter-twin cooperation}

Fig.~\ref{fig:DT_lifecycle} shows an overview of the wide variety of applications and environments that DT-supervised SGs are envisioned to support and a high-level description of DT's main functionalities. 
In local DTs, synchronous communication between digital and physical twins is expected to mimic the actual system layout and record the SG time-synchronized data. On the other hand, global-view DTs use asynchronous communication to import distributed information about the system's parts and export its computed model weights to other DTs. 
Seamless connectivity between the cyber twins is required to maintain the load-supply gap and respond to generation output shortages, discrepancies from nominal grid frequency, and sudden load shedding, among other things.
 
Hence, inter-twin coordination is required for these types of sensitive tasks to ensure integrated resource management at all times. 
Graph neural networks enable DTs to formulate intricate relationships between real entities, which helps them to model the monitored physical objects. Moreover, matching theory provides relational connections in the digital plane between individual DTs for better cooperation and functional coordination. 
This interplay between the DTs can set in motion a continuous and developmental process that immunizes the SG against unexpected events. Hence, the DTs that supervise substations that have experienced outage events in the past can share knowledge with other DTs whose substations have not experienced similar incidents.
Resource allocation can be performed more efficiently with the broader view that DTs gain from their interactions with each other.

\subsection{Different twin technologies}

Enabling communication technologies must be investigated to satisfy DT networking requirements.
In a heterogeneous SG that is influenced by different factors, e.g. generation and consumption gap balance, DER integration, and intelligent transportation, it is natural for DTs to require multiple technologies that meet their needs.
Real-time monitoring DTs that require high throughput for their short-range applications should use cellular $5$G technology.
DTs that supervise IoV scenarios need to use long-range communication technologies that support mobility and include sustainable handover strategies.
Micro DTs that monitor IIoT networks are best to use cellular narrow-band IoT (NB-IoT) technology in which the bulk of communications is massive machine-type communications. On the other hand, the long-term evolution category M$1$ (LTE-M) is best suited for mission-critical communications that require ultra-low-latency data transmission. 

Model sharing and model propagation tasks are assumed to require a low data rate with sporadic transmission
and be delay-insensitive, as compared to monitoring applications.
Therefore, a DT infrequently reports changes in its model weights and receives updated model insights from other DTs.
License-free technologies such as Long-Range (LoRa) and SigFox are suitable for model sharing in inter-DT communication. SigFox has a longer transmission range, while the data rates for both technologies are variable, with relaxed delay requirements. Cognitive radio technologies, e.g. wireless regional area network (WRAN), can provide 
a longer coverage range when needed, i.e. between DT networks that are geographically far apart. Reliable routing algorithms can also propagate the DTs' model weights across the network.
For event response operations, DT networks should support high data rates and ultra-low latency. Although unexpected events might occur rarely, their corresponding DT operations should be served by dedicated networks. Low Earth orbit (LEO) satellite communications are suitable as they 
provide infrastructure independence, universal coverage, resilience against adverse weather conditions and broad range service, which makes them ideal for SG event response tasks and rescue operations.

\section{Digital Twinning for Proactive Smart Grids}

%Future smart grid will depend on individual producers and consumers of renewable energy resources such as wind and solar, making versatile trading between each other. DT would be useful by offering novel strategies in how distributed trading can be performed by firstly obtaining information about the customer needs and producer goods. DT-based energy trading architectures could optimally match customers and producers based on availability, cost and trade history etc. Blockchain and underlying smart contracts could aid in securing the trading schemes.

\begin{figure*}
	\centering
	\includegraphics[width=16cm]{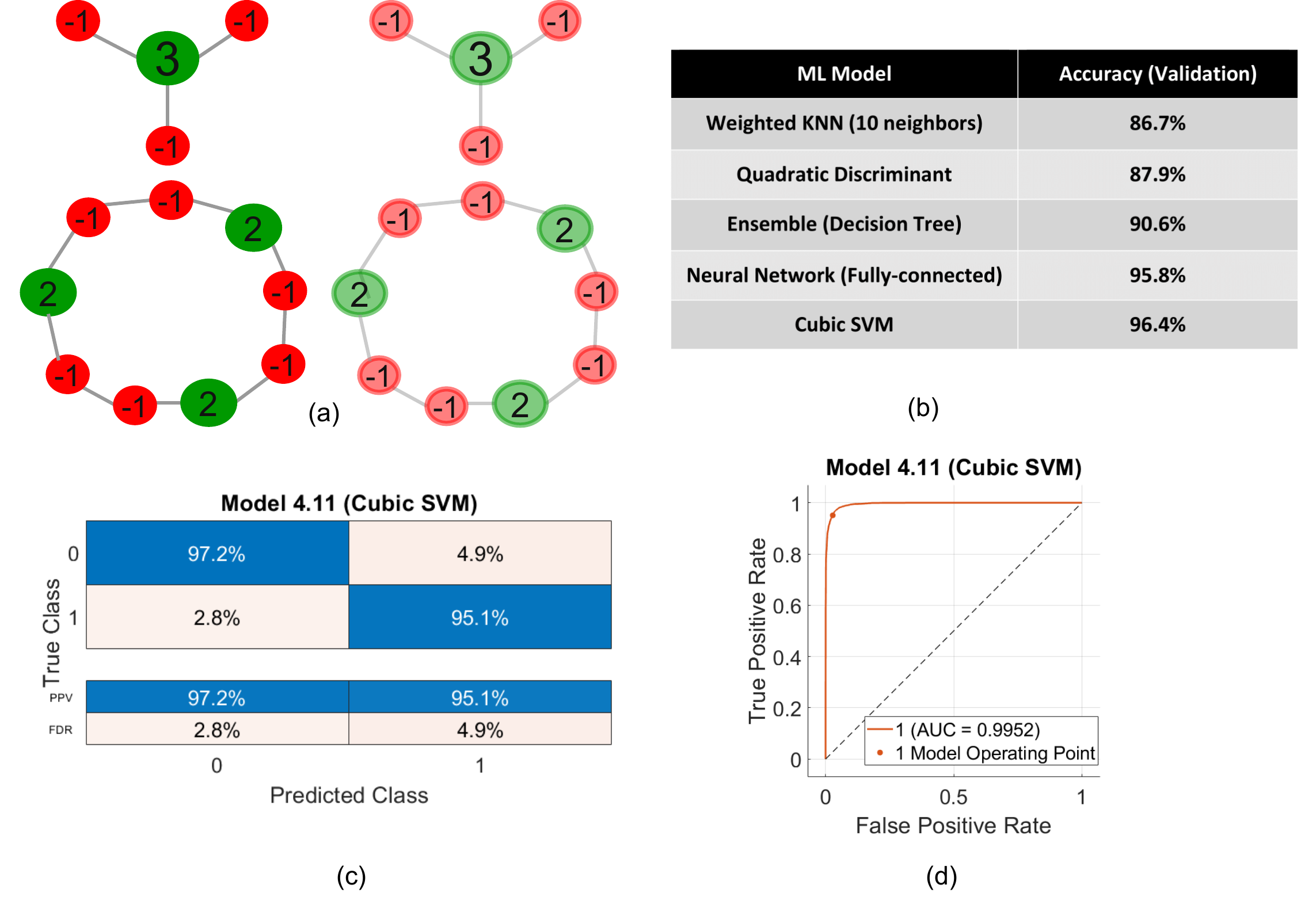}\\
	\caption{$4-$ and $9-$node grids based on \cite{Schafer2016} are shown in (a) along with their initial DT models, where green and red nodes denote producers and consumers, respectively. A synthetic dataset of $10,000$ labeled examples about specific parameters of the grid nodes is used by the DT to train ML models \cite{Arzamasov2018}, to classify whether the monitored grid is stable or not. The performance of these classifiers is shown in (b), where Support Vector Machines with Cubic kernel have presented the highest accuracy on validation sets. The confusion matrix is shown in (c), with results for positive predicting values and false discovery rate. The receiver operating characteristics curve is shown in (d) for positive decision (i.e. grid is unstable).}
	\label{fig:DT_example}
\end{figure*}

\subsection{Event forecasting and prevention}

Unexpected malfunctions are common in power systems and can result in service interruptions. 
A DT-monitored SG regularly takes steps to anticipate such situations by ensuring a balance between energy supply and demand. 
More specifically, phasor measurement units provide high precision time-stamped data, enabling the DTs to visualize the grid's internal mechanisms, locate sources of trouble, and isolate them if necessary using circuit breakers, for example. 
Proactive resource allocation can be customized and incorporated into DT software to anticipate and prevent anomalous events in the SG. Weather data, load forecasts, DER analysis, can all be employed by the DT to preclude preventable outages by detecting and isolating fragile parts of the grid that could bring down the system. 

As the SG is going to integrate DERs to reduce carbon footprint, DT-empowered communications can be a key-enabler to design network optimization tools, and debug and troubleshoot when needed \cite{Khan2022}. 
A demand-side management example of grid-decentralized control that assists with DERs' integration is shown in Fig.~\ref{fig:DT_example}. This application uses a variable price mechanism to motivate consumers to adapt their consumption behavior to regulate the SG's frequency \cite{Schafer2016}. The DTs use different ML tools to train a classifier fed with certain information about the grid nodes and decides whether the monitored grid is stable or unstable. Then, of course, this information would be used to take further action, such as unit commitment or load shedding, in order to restore normal operation if the grid is found to be unstable.

%Microgrids are key component of future smart grids, where a microgrid represents a semi-independent entity that can perform energy trading with the central grid and with each others; therefore alleviating potential blackouts by supplying stored power especially during high demand times such as rush hours. A DT-based scheme that captures real-time and history information about the energy needs and capabilities of microgrids and central grid could help to find the best resource optimization forecasting decisions to avoid destabilizing phenomena such as blackouts and power shortages during summer season in the middle east or winter storms in Canada, where demand on cooling/heating is at highest levels.

%\hl{How can RRM-based on AI tools and model-driven tools address these problems?

%<Proactive RRM-specific DTs>
%- what is needed for RRM to grow to match DT capabilities? Very big question to envision and imagine.
%One possibility is to prevent future outages by invoking data forecasting and prediction methods to analyze SG status given huge versatile
%data about weather, load, information about weak points of the system, and mission-critical users. Hypothetical scenarios can be imagined
%using proactive RRM to study future events and choose best possible reaction given different scenarios. Available data about past outages
%and events can help to make the scenarios realistic by hypothesizing a similar order of magnitude and effect on the system.}

%, where the area under the curve (AUC) is $0.9952$

\subsection{Twin-assisted recovery and support}

Disasters could disrupt communication infrastructure in affected regions, which impacts coverage service for these areas. 
Communication link failures can happen in disasters, and DTs should swiftly activate aftermath response.
During disasters and grid response, data processing can cause additional load on the already strained local DT network. Hence, it is imperative that computation offloading to other DTs can be leveraged to ease the burden on constrained resources. However, since this increases the load on these DTs, prioritizing schemes can be employed to favor critical data types and ensure maximum resource utilization.
An elastic response would be essential to protect connected loads from being affected by a widespread outage event. 

In addition, constrained load-frequency control is essential to balance the grid and start the stable recovery of services in affected regions. 
DT-based resource allocation must be on-demand and use optimization tools to provide automated assistance and seek the optimal service restoration actions to supply the physical twin system. DTs will test the automated protection mechanisms under extreme conditions without compromising the integrity of the physical grid.
Moreover, numerous protective relays within harsh SG environments are prone to frequent failures, and virtual monitoring of DTs can help in scheduling and performing predictive maintenance, in addition to verifying test results and detecting defective equipment, leading to timely part replacement and uninterrupted service.
Automated dynamic DT resource management is necessary for disaster management, taking into account the time-varying nature of wireless DT networks and reliable information dissemination requirements. At the same time, DT-supported response should meet stringent task completion latency and avoid overloading the grid's network resources.

%When unforeseen emergencies occur in DT-tied SG, recovery often reacts in local information paradigm. In this case, the system reacts in slow motion, often having a ripple effect based on scale level of the inflicted damage. However, from DT perspective, local DT first assesses the distress based on its communication with the physical system. It amasses the right volume of assistance, and shares the news with neighboring DTs. If the service disruption is wide-spread for example, other DTs can jump in and share some of the burden. They can -for example- lend some of available resources to the affected areas. Optimization would be done to favor vital infrastructure with critical priorities. DTs could do the post-accident recovery and assistance automatically, based on AI-optimization to seek the optimal actions given the physical twins states.

\subsection{Smart grid resilience and auto-correction}

A resilient SG must be self-sustaining and self-correcting.
Microgrids, for example, should be community-resilient to deal with possible outage events while providing essential support for primary critical services. 
Network layer protocols will be designed by the DT per need to respond to ad-hoc network architectures where device-to-device communications can be used.
DT-supported predictive maintenance is necessary for the early detection of any vulnerable points in the grid. For example, in an automated substation, protective relaying can be monitored and periodically tested by DTs to ensure that over-voltage protection relays meet applicable minimum standards. 
The voltage disturbance correction feature must always respond when potential transformers report a power imbalance. 

Future DT-based resource management should answer questions such as how to send and route assistance to distressed regions from neighboring areas. Coordination plans enable macro DTs, with their global-view advantage, to assign available resources to alleviate stress in concerned areas. Micro DTs should use precise information to work on a real-time frame to provide resources and post-event assistance until regular service is restored.

%Security is a major issue in smart grid, which can be susceptible to attacks of various sorts. One example where DT can help with security is to build a virtual simulator at the DT layer that simulates cyber attacks on the network and find the weak points, also known as ethical hacking.
%Next, this information can be used to strengthen the security of the blockchain system. However, some challenges that faces DT-based blockchain security scheme are intensive resource consumption and high latency. These are challenges that should be addressed for any future DT blockchain models.

%\hl{<Mission-critical and non-critical services>
%- Hosptials and community essential services such as firefighting should always have power, especially when a natural disaster occurs.
%These services would always be prioritized in any scenario, and this would also be reflected in the DT modeling and RRM algorithms.
%For this purpose, when an area is hit, such as a MG, it is important to route available help from other unaffected areas with minimal delay
%and efficiency. This is where macro DTs can anticipate and work immediately following unseen events to override local DTs and use available
%resources to alleviate stress on the affected areas. Algorithms and methods are needed to find harmony between the two layers DTs to 
%work on miniscule and majscule time frames for the everyday RRM services and proactive crisis-preventative and post-crisis driven RRM
%tools.}

\section{Challenges and Future Directions}

\subsection{Advanced communication technologies}

The development of beyond-$5$G technologies can help enable faster and more reliable communication between DTs and other SG components. The ultra-reliability of transmitted data must be ensured to maintain continuous independence and mirroring between the digital and physical planes. 
Leveraging edge computing to process data closer to the source can help improve the real-time performance of DTs.

\subsection{Model Complexity and Scalability}

Developing and managing complex DT models for the entire SG can be challenging and require significant computational resources. With millions of physical components in SGs, explicitly representing them in precise detail while sustaining the evolving granularity of DT models remains a crucial challenge for current computational architectures. 
Moreover, separating old and new sensed data is crucial for DT model development, where the freshness of monitored data must be tracked using metrics like the age of information.

\subsection{Interoperability}

Heterogeneous DTs that deal with 
	different SG use cases should have a compatible message structure to facilitate information sharing and ensure reliable inter-DT communication. Interoperability between DT and other SG technologies is essential for seamless communication and data exchange. However, integrating DTs with existing legacy systems and infrastructure can be challenging due to nonstandard and non-interoperable communication protocols.
%On the other hand, constructing and update of DT models needs to be done in congruence with other DTs. This experience driven approach allows the DTs with limited local data to acquire global information through bi-directional communication with neighboring DTs. Federated learning and blockchain can help in propagating the local information securely and efficiently while local DTs always refresh their models. This slow spread of model parameters with continuous update immunizes the DTs against unexpected events that could occur due to blissful ignorance of certain DTs who experience a common performance in the physical twins. Other DTs who experience anomalous pattern in the associated physical twins help their fellow DTs get heads up about what kind of operations can happen in the real-system.

%\hl{- Model aggregation between DTs: After each DT constructs its own model of the underlying physical system, and keeps updating it, these
%models can be used in blockchain fashion to construct global models using macro DTs, that will aggregate the layer 2 DT model parameters.
%The global model is not a sum of local models, but needs to have larger time frame, it should look for more recurrent themes, and anticipate
%long-term issues or benefits. The wide-system view of macro DTs should consider also inter-sharing between each other. Macro DTs can work
%to anticipate problems and work in preventative manner to allocate the resources more efficiently, by developing game-theoretic models
%of underlying DTs and physical twins, and finding Nash equilibrium points.}

\subsection{Big data quality, privacy and security}

DTs rely heavily on data, and storage and processing capabilities are indispensable for handling this data. Privacy and security issues could arise for pervasive DTs when they start to collect SG users' behavior patterns and due to malignant intruders. The use of blockchain technology and federated learning to enhance the security, privacy, and interoperability of DTs is a promising area of research. However, latency incurred by blockchain technologies for mission-critical applications and real-time monitoring in SG needs to be addressed.

\section{Conclusion}

% Monitoring DTs can retain two models at any time---a short-term one and a long-term one. The former can be adjusted occasionally based on insights and lessons that improbable events have contributed to the DTs' long-time view. However, the smart grid's long-term sustainability depends on a macro-scale model that takes into account larger information gathered over an extended period of the DTs' lifetime. 

Digital twin is an emerging technology that is anticipated to continue increasing in popularity and gain more recognition and attention in industrial and academic circles. In this article, a principle-centered overview of a digital twin-supported smart grid is presented and followed by a dissection of how digital twins can immunize the smart grid against likely threats and ensure its resource management tasks are performed more effectively in a variety of smart grid use cases. We conclude by outlining a few challenges to be aware of with regard to digital twins and prospective future research directions for further consideration.

\bibliographystyle{ieeetr}
\bibliography{library}

\section*{Biographies}
\small
\noindent\textbf{Abdullah Othman} is currently pursuing the Ph.D. degree at the École de technologie supérieure (ÉTS), Université du Québec, Montréal, Canada. His research interests include wireless communications, smart grid, and digital twins.
\\ \\
\textbf{Georges Kaddoum} is a Professor and Canada Research Chair with the ÉTS. His recent research activities cover wireless communication networks, tactical communications, resource allocations, and security. \\ \\
\textbf{Joao V. C. Evangelista} received his Ph.D. from the ÉTS and is currently with Ericsson Canada. His current research interests include machine-to-machine communications, non-orthogonal multiple access, and stochastic geometric modeling.\\ \\
\textbf{Minh Au} is currently a Researcher Scientist with Hydro-Quebec
Research Institute (IREQ), Varennes, QC, Canada. His research interests include digital transformation of power substation, partial discharge phenomenon,  information theory
and cyber-security for smart grid.\\ \\
\textbf{Basile L. Agba} is currently Manager of Vision and Strategies at IREQ and Adjunct Prof. at the ÉTS. His research interests include channel
modeling in high voltage environments, smart grid communications,
substation automation, and cyber-security.

%\end{thebibliography}

\end{document}